\documentclass[12pt]{article}
\usepackage{graphicx}
\usepackage{subfigure}

\def\EtMiss  {E_\mathrm{T}\!\!\!\!\!\!\!/ \ \; }

\newcommand\pubdate{\today}

\def\wuppertal{Bergische Universit\"at Wuppertal \\ 
 Gau{\ss}stra{\ss}e 20, D-42119 Wuppertal, GERMANY}
\def\support{\footnote{Work supported by the Helmholtz Alliance 
            ``Physics at the Terascale''.}}

\def\Title#1{\begin{center} {\Large #1 } \end{center}}
\def\Author#1{\begin{center}{ \sc #1} \end{center}}
\def\Address#1{\begin{center}{ \it #1} \end{center}}

\newcommand\pubblock{\rightline{\begin{tabular}{l} 
\\
         \pubdate  \end{tabular}}}
\newenvironment{Abstract}{\begin{quotation}  }{\end{quotation}}
\newenvironment{Presented}{\begin{quotation} \begin{center} 
             PRESENTED AT\end{center}\bigskip 
      \begin{center}\begin{large}}{\end{large}\end{center} \end{quotation}}
\def\Acknowledgements{\bigskip  \bigskip \begin{center} \begin{large}
             \bf ACKNOWLEDGEMENTS \end{large}\end{center}}
\input econfmacros.def

\begin{document}
\begin{titlepage}
\pubblock

\vfill
\Title{Direct Measurements of $V_{tb}$}
\vfill
\Author{ Wolfgang Wagner\support}
\Address{\wuppertal}
\vfill
\begin{Abstract}
The measurement of the single top-quark production cross section in 
hadron collisions at the Tevatron and the LHC can be used to determine 
the absolute value of the CKM matrix element $V_{tb}$ without assuming 
the unitarity of the CKM matrix. By measuring the branching ratio
of the decay $t\rightarrow W + b$ one can relate $V_{tb}$ to the
matrix elements $V_{ts}$ and $V_{td}$. The current experimental status and
future prospects of these measurements are reviewed.
\end{Abstract}
\vfill
\begin{Presented}
CKM 2010 \\
the $6^\mathrm{th}$ International Workshop on the CKM Unitarity Triangle \\
University of Warwick, United Kingdom, September 6--10, 2010
\end{Presented}
\vfill
\end{titlepage}
\def\thefootnote{\fnsymbol{footnote}}
\setcounter{footnote}{0}
\section{Introduction}
\begin{figure}[t]
    \begin{center}
    \subfigure[]{
    \includegraphics[width=0.18\textwidth]{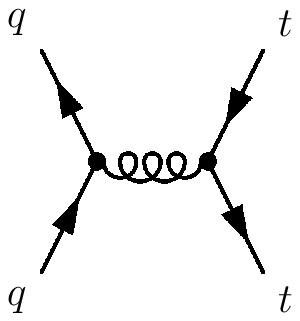}
      \label{subfig:qqtt}}  
    \hspace*{16mm}
    \subfigure[]{
    \includegraphics[width=0.18\textwidth]{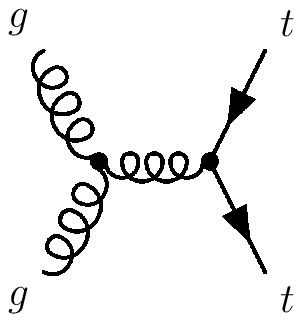}  \hspace*{4mm}
    \includegraphics[width=0.18\textwidth]{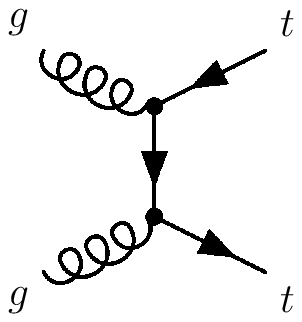}  \hspace*{4mm}
    \includegraphics[width=0.18\textwidth]{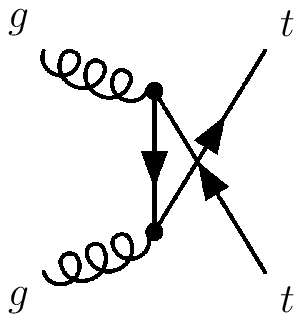}
      \label{subfig:ggtt}}
    \end{center}
    \caption{\label{fig:leadingOrderttbar} Feynman diagrams of the leading order
      processes for $t\bar{t}$ production: 
      \subref{subfig:qqtt} quark-antiquark annihilation 
      ($q\bar{q} \rightarrow t\bar{t}$) and \subref{subfig:ggtt}
      gluon-gluon fusion ($gg \rightarrow t\bar{t}$).
     }
  \end{figure}

The main source of top quarks at the Tevatron is the pair production
via the strong interaction. At leading order in perturbation theory 
($\alpha_s^2$) there are two processes that contribute to $t\bar{t}$ production,
quark-antiquark annihilation $q\bar{q} \rightarrow t\bar{t}$ and 
gluon-gluon fusion $gg \rightarrow t\bar{t}$.
The corresponding Feynman diagrams for these processes are depicted
in Fig.~\ref{fig:leadingOrderttbar}.
At the Tevatron the $t\bar{t}$ cross section is dominated by $q\bar{q}$ 
annihilation processes contributing 85\%, while only 15\% of 
$t\bar{t}$ pairs are produced via $gg$ fusion. At the LHC at 
$\sqrt{s}=14\,\mathrm{TeV}$, the situation
is reversed, 90\% of $t\bar{t}$ events are produced by $gg$ fusion 
processes and only 10\% by $q\bar{q}$ annihilation. The reason for this 
phenomenon is the very different parton luminosities at partonic energies 
above the $t\bar{t}$ threshold. Due to the higher center of mass energy 
at the LHC it is possible to produce $t\bar{t}$ pairs already at lower
$x$ of the incoming partons where the gluon parton density dominates over
the quark densities. In addition, the Tevatron features antiquarks as
constituent quarks of the antiproton leading to a considerable large 
antiquark density at large $x$. The cross sections for
$t\bar{t}$ production are predicted \cite{Langenfeld:2009wd} to be
\begin{eqnarray}
\sigma_\mathrm{LHC} & = & (887^{+9}_{-33}\,(\mathrm{scale})
^{+15}_{-15}\,(\mathrm{PDF}))\;\mathrm{pb} 
 \quad(\mathrm{at} \ 14\;\mathrm{TeV}) \, , \\
\label{eq:physics:top:ttbarxs_theory}
\sigma_\mathrm{Tev} & = & (7.04^{+0.24}_{-0.36}\,(\mathrm{scale})
^{+0.14}_{-0.14}\,(\mathrm{PDF}))\;\mathrm{pb} 
\quad(\mathrm{at} \ 1.96\;\mathrm{TeV})\, .
\end{eqnarray}
According to the standard model (SM) top
quarks decay with a branching ratio of nearly 100\% to a bottom quark
and a $W$ boson and the $t\bar{t}$ final states can be 
classified according to the decay modes of the $W$ bosons. The most
important (or golden) channel is the so-called {\it lepton+jets} channel
where one $W$ boson decays leptonically into a charged lepton (electron or 
muon) plus
a neutrino, while the second $W$ boson decays into jets.
The lepton+jets channel features a large branching ratio
of about 29\%, manageable backgrounds, and allows for the full 
reconstruction of the event kinematics. 
Other accessible channels are the {\it dilepton}
channel, in which both $W$ bosons decay to leptons, and the {\it all-hadronic}
channel, where both $W$ bosons decay hadronically. 
The dilepton channel has the advantage of having a low background, 
but suffers on the other hand from a lower branching fraction (5\%) 
compared to the lepton+jets channel.
The all-hadronic channel, on the contrary, has the largest branching 
ratio of all $t\bar{t}$ event categories (46\%), but 
has the drawback of a huge QCD multijet background, that has to be
controlled experimentally. The different categories of $t\bar{t}$
and their branching fractions are summarized in 
Table~\ref{tab:ttbarDecay}.
\begin{table}
  \begin{center}
  \begin{tabular}{@{}cccc@{}}
    \hline
    $W$ decays & $e/\mu\nu$ & $\tau\nu$ & $q\bar{q}$ \\ \hline
    $e/\mu\nu$ &   5\%     &   5\%     &    29\%    \\
    $\tau\nu$  &   --       &  1\%     &    15\%    \\
    $q\bar{q}$ &   --       &   --      &   46\%    \\ \hline
  \end{tabular} 
  \end{center}
  \caption{\label{tab:ttbarDecay} Categories of $t\bar{t}$ events and their 
    branching fractions. The sum of all fractions is above 100\% because
    of rounding effects.}
\end{table}

\section{Top-antitop Cross Section Measurements}
The most precise $t\bar{t}$ cross section measurements are obtained in the 
lepton+jets
channel based on analyses in which the background is controlled either 
by the identification
of the $b$-quark jets or by exploiting kinematic or topological
information, see for example~\cite{Abazov:2008gc}.
The experimental signature of lepton+jets $t\bar{t}$ events
comprises a reconstructed isolated lepton candidate, large
missing transverse energy ($\EtMiss$) and at least four jets with large
transverse energy $E_T \equiv E\cdot\sin \theta$. Two jets originate
from $b$-quarks. Typical selection cuts ask for a charged lepton with 
$p_T > 20\;\mathrm{GeV}/c$, $\EtMiss > 20\;\mathrm{GeV}$, and at least
four jets with $E_T > 20\;\mathrm{GeV}$ and $|\eta|<2.0$, one of them
identified as a $b$-quark jet. The most commonly used algorithm to 
identify $b$-quark jets is based on the reconstruction of secondary
vertices in jets, exploiting the relatively long lifetime of $b$-hadrons
and a large Lorentz boost. The typical decay length of $b$-hadrons
in high-$p_T$ $b$-quark jets is on the order of a few millimeters.
The requirement of a secondary vertex within one of the jets leads to
a large reduction of the $W$+jets background by roughly a factor
of 50, while the selection efficiency for $t\bar{t}$ events is 
about 50\% to 60\%.

\begin{figure}[t]
\begin{center}
\includegraphics[width=0.5\textwidth]{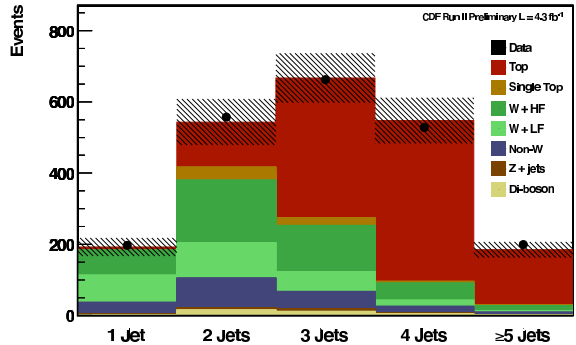}
\end{center}
\caption{\label{fig:Wjets}Jet multiplicity distribution for the 
  $W$+jets data set, where the $W$ boson is reconstructed in its leptonic 
  decay
  $W^\pm \rightarrow \ell^\pm \nu_\ell (\bar{\nu}_\ell)$. A cut on 
  $H_T>230\,\mathrm{GeV}$ was applied. The analyzed data set corresponds
  to $4.3\,\mathrm{fb^{-1}}$ of CDF data.} 
\end{figure}

The most recent CDF analysis based on secondary vertex 
$b$ tagging~\cite{Aaltonen:2010ic} 
is a counting experiment in which the background rate is estimated using a combination 
of simulated events and data driven methods. The signal region is
defined as the data set with a leptonic $W$ candidate plus $\geq 3$
jets. To further suppress background, a cut on the sum of all transverse
energies $H_T>230\,\mathrm{GeV}$ is applied. 
The jet multiplicity distribution of the $W$+jets data set observed by
this CDF analysis is shown in Figure~\ref{fig:Wjets}.
The uncertainty on the luminosity measurement is reduced by measuring
the ratio of $t\bar{t}$-to-$Z$-boson cross sections and the measured
cross section is found to be
$7.32\pm 0.36\,(\mathrm{stat})\pm 0.59 (\mathrm{syst})\pm 0.14\,(\mathrm{Z\;theory})\;\mathrm{pb}$, assuming $m_t = 172.5\;\mathrm{GeV}/c^2$. 
The cross section of $t\bar{t}$ production and  
$Z/\gamma^* \rightarrow \ell^+\ell-$ production are measured in 
data samples corresponding to the same integrated luminosity.
By forming the ratio of both measured cross sections and multiplying
by the well-known theoretical $Z/\gamma^* \rightarrow \ell^+\ell-$
cross section the luminosity uncertainty of 6\% is effectively removed
and replaced by the uncertainty on the  $Z/\gamma^*$ cross section
of 2\%. In a similar measurement the D\O collaboration measured
$\sigma_{t\bar{t}} = 7.93 ^{+1.04}_{-0.91}\;\mathrm{pb}$~\cite{D0_CONF_6037}.

Recently, the first observations of $t\bar{t}$ pairs produced 
at the LHC have been made by the ATLAS and CMS 
collaborations~\cite{Collaboration:2010ey,Khachatryan:2010ez}.
The measured $t\bar{t}$ cross section at $\sqrt{s}=7\,\mathrm{TeV}$ is
$\sigma_{t\bar{t}} = 145 \pm 31\, (\mathrm{stat})\; 
 ^{+42}_{-27}\,(\mathrm{syst})\;\mathrm{pb}$
and 
$\sigma_{t\bar{t}} = 194 \pm 72\, (\mathrm{stat}) \pm 24\,(\mathrm{syst})
 \pm 21 (\mathrm{lumi})\;\mathrm{pb}$, respectively.

\section{The measurement of $\mathcal{R}_b$}
\label{subsec:top_Rb}
In the SM the top quark is predicted to decay to a $W$ boson and a $b$ quark
with a branching fraction 
$\mathcal{R}_b \equiv \mathrm{BR}(t\rightarrow Wb)$ close to 100\%.
This prediction is obtained in the following way.
In general, the top quark can decay in three channels 
$t\rightarrow d/s/b +W^+$ and $\mathcal{R}_b$ is given by the ratio of 
the squares of the relevant CKM matrix elements: 
$\mathcal{R}_b = |V_{tb}|^2/(|V_{td}|^2+|V_{ts}|^2+|V_{tb}|^2$).
In the SM the CKM matrix has to be unitary 
($\mathbf{V V^\dag} = \mathbf{V^\dag V} = \mathbf{1}$), which 
leads to $|V_{td}|^2+|V_{ts}|^2+|V_{tb}|^2 = 1$ and thereby to
$\mathcal{R}_b = |V_{tb}|^2$.
Our present knowledge on $|V_{tb}|$ stems primarily from measurements 
of $b$-meson and $c$-meson decays which determine the values of the 
other CKM matrix elements. Using the unitarity condition of the CKM matrix 
one can obtain $|V_{tb}|$ in an indirect way. 
This method yields
$|V_{tb}| = 0.999133 \pm 0.000044$ with very high precision~\cite{Amsler:2008zzb}. 

Only recently a determination of $|V_{tb}|$ without unitarity assumption
was obtained from the measurement of the single top-quark cross section,
yielding $|V_{tb}|=0.88\pm0.07$~\cite{Group:2009qk}.
However, if a fourth generation of quarks was present, the unitarity of 
the $3\times 3$ CKM matrix could be violated. Therefore, it is desirable to 
make a direct measurement of $\mathcal{R}_b$ using $t\bar{t}$ candidate events.

In most $t\bar{t}$ cross section analyses the assumption $\mathcal{R}_b=1$ is 
made, but CDF and D\O \ have also made two measurements without this 
constraint~\cite{Affolder:2000xb,Acosta:2005hr,Abazov:2008yn}. 
In the latest analysis from D\O~\cite{Abazov:2008yn} the $W$+jets data set is split 
in various disjoint 
subsets according to the number
of jets (0, 1, or $\geq 2$), the charged lepton type (electron or muon),
and most importantly the number of $b$-tagged jets. The fit results are:
$\mathcal{R}_b = 0.97^{+0.09}_{-0.08}$ and 
$\sigma(t\bar{t}) = 8.18^{+0.90}_{-0.84} \pm 0.50\;(\mathrm{lumi})\,\mathrm{pb}$,
where the statistical and systematic uncertainties have been combined.
The lower limit on $\mathcal{R}_b$ is determined to be 
$\mathcal{R}_b > 0.79$ at the 95\% C.L. In terms of the CKM matrix 
elements this result can be summarized by the following relation:
\begin{equation}
 |V_{td}|^2 + |V_{ts}|^2 < 0.263 \cdot |V_{tb}|^2
\end{equation}
at the 95\% C.L. The measurement quoted above is based on a data
set corresponding to $0.9\,\mathrm{fb^{-1}}$, which is only a small 
fraction of the available data at the Tevatron. In this measurement
the statistical and systematic uncertainties have about the same size.
If the an update to the full data set was made, the measurement would 
be completely systematically limited.

\section{Measurement of Single-Top Production}
\begin{figure}[!h!tpb]
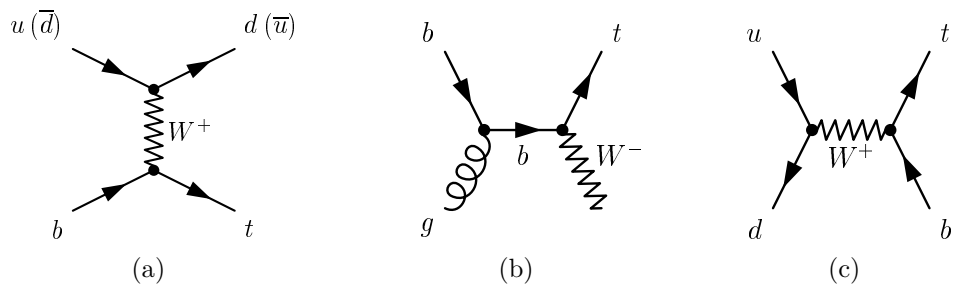

  \centering
  \subfigure[]{
    \includegraphics[width=0.25\textwidth]{singTopWg3Pict.epsi}
    \label{fig:tchan}
  }
  \hspace*{0.07\textwidth}
  \subfigure[]{
    \includegraphics[width=0.18\textwidth]{singTopAssocPict.epsi}
    \label{fig:Wt}
  }
  \hspace*{0.07\textwidth}
  \subfigure[]{
    \includegraphics[width=0.18\textwidth]{singTopSChanPict.epsi}
    \label{fig:schan}
  } 
\caption{Feynman diagrams of single top-quark production processes.
         \subref{fig:tchan} $t$ channel production, 
         \subref{fig:Wt} associcated $Wt$ production, and
         \subref{fig:schan} $s$ channel production.}
\label{fig:feynman}
\end{figure}

While $t\bar{t}$ pair production via the strong interaction is the main source
of top quarks at the Tevatron and the LHC, top quarks can also be produced
singly via weak interactions involving the $Wtb$ vertex. There are three
production modes which are distinguished by the virtuality, $Q^2$, 
of the $W$ boson.
The dominant 
process is the $t$ channel exchange of a virtual $W$, depicted in 
Fig.~\ref{fig:feynman}\subref{fig:tchan}.
At the LHC, the subprocess with the second highest cross section 
is the associcated production of an on-shell $W$ boson and a 
top quark, see Fig.~\ref{fig:feynman}\subref{fig:Wt}. 
At the Tevatron, $Wt$ production has a negligible cross section
compared to the other two processes.
The Drell-Yan type production
of a $t\bar{b}$, see Fig.~\ref{fig:feynman}\subref{fig:schan}, 
has only subleading character at the LHC, but is relevant at
the Tevatron.

While Run~1 (1992--1996) and early Run~2 searches for single top quarks at
CDF and D\O \ could only set upper limits on the production cross
section~\cite{Acosta:2004bs,Abazov:2006uq}, analyses using more data and
advanced analysis techniques were able to observe
singly produced top quarks in March 2009 with a significance of
five Gaussian standard 
deviations~\cite{Aaltonen:2009jj,Aaltonen:2010jr,Abazov:2009ii}.
These analyses consider both production modes relevant at the Tevatron,
$t$ channel and $s$ channel, as one single-top signal, assuming the ratio of
$t$ channel to $s$ channel events to be given by the SM. This search strategy
is often referred to as {\em combined search}. By measuring the inclusive
single top-quark cross section and using theory predictions, one can
deduce the absolute value of the CKM matrix element $V_{tb}$, without
the assumption that there are only three generations of quarks. If one measured $|V_{tb}|$
to be significantly smaller than one this would be a strong indication for a
fourth generation of quarks or other effects beyond the
SM~\cite{Eberhardt:2010bm,Bobrowski:2009ng,Alwall:2006bx}.

Single top-quark events feature a $Wb\bar{b}$ ($s$-channel) or 
$Wbq\bar{b}$ ($t$-channel) partonic final state.
The most sensitive analyses reconstruct the $W$ boson originating from the
top-quark decay in its leptonic decay modes $e\nu_e$ or $\mu\nu_\mu$,
while hadronic $W$ decays and decays to $\tau\nu_\tau$ are not explicitly
considered, because of large backgrounds from QCD-induced jet production.
The quarks from the hard scattering process manifest themselves as
hadronic jets with large transverse momentum. Additional jets may arise 
from hard gluon radiation in the initial or final state.
The experimental signature of SM single top-quarks is therefore given
by one isolated high-$p_T$ charged lepton (electron or muon),
large missing transverse energy ($\EtMiss$), and two or three 
high-$E_T$ jets,
of which one or two originate from a $b$ or $\bar{b}$ quark.

The dominating background process is $Wb\bar{b}$, followed by misidentified
$W +$ light-quark jet events, $Wc\bar{c}$, and $Wcj$. 
In the $W+3$ jets data-set $t\bar{t}$ is the most important background.
The total event detection efficiency of single top-quark events
is about 2\% for the $t$-channel process and about 3\% for
the $s$ channel.
Even though the single top-quark production cross section is
predicted to amount to about 40\% of the $t\bar{t}$ cross section,
the signal has been obscured for a long time by the very challenging
background.
After the cut-based event selection sketched above,
the signal-to-background ratio is only about 5 to 6\%.
Further kinematic cuts on the event topology
proved to be prohibitive, since the number of signal events in the
resulting data set would become too small. Facing this challenge,
the analysis groups in both Tevatron collaborations turned to
multivariate techniques, in order to exploit as much information
about the observed events as possible. The explored techniques comprise
artificial neural networks, LO matrix elements,
boosted decision trees, and likelihood ratios.
All these techniques combine the information contained in several
variables into one powerful discriminant, maximising the
separation between the single top-quark signal and the backgrounds.
All multivariate analyses see a signal of single top-quark production
with significances ranging from 2.4 to 5.2 Gaussian standard deviations.
To obtain the most precise result, all analyses of the two collaborations
are combined, resulting in a single top-quark cross section of
$2.76^{+0.58}_{-0.47}\;\mathrm{pb}$
(at $m_t= 170\,\mathrm{GeV}/c^2$)~\cite{Group:2009qk}.

The measured single top-quark production cross sections can be used to
determine the absolute value of the CKM-matrix element $V_{tb}$, if one assumes
$V_{tb} \gg V_{ts}$, $V_{tb} \gg V_{td}$, and a SM-like left-handed
coupling at the $Wtb$ vertex.
Contrary to indirect determinations
in the framework of flavour physics the extraction
of $|V_{tb}|$ via single top-quark production does not assume unitarity
of the CKM matrix and is thereby sensitive to a fourth generation
of quarks. The assumption of $V_{ts}$ and $V_{td}$ being small compared
to $V_{tb}$ enters on the production side, since top quarks can also be
produced by $Wts$ and $Wtd$ vertices, and in top-quark decay.
The phenomenological analysis of~\cite{Eberhardt:2010bm,Bobrowski:2009ng} 
as well as the measurement of $R_b$, see~\ref{subsec:top_Rb}, indicate
that this assumption is well justified.
To determine $|V_{tb}|$ the analysts divide the measured single top-quark
cross section by the predicted value, which assumes $|V_{tb}|=1$, and take
the square root.
Based on the combined cross-section result the Tevatron
collaborations obtain $|V_{tb}|=0.88\pm0.07$~\cite{Group:2009qk}.

\section{Conclusion}
Measurements of top-quark production and decay add valuable 
information to determine or constrain the CKM matrix elements 
$V_{tb}$, $V_{ts}$, and $V_{td}$. These experimental constraints
are complementary to measurements made with $b$ and $c$ hadrons.
The measurements of $R_b$ and the single top-quark cross section
lead currently to the contraints
\[
 |V_{td}|^2 + |V_{ts}|^2 < 0.263 \cdot |V_{tb}|^2 \ \ 
 \mathrm{at \ the} \ 95\%\,\mathrm{C.L.} \ \mathrm{and}
\]
\[ |V_{tb}|=0.88\pm0.07\;. \]
In the future it may be possible to also constrain $V_{ts}$ and 
$V_{td}$ in measurements of single top-quark production, even 
though it will be very challenging to disentangle the different 
production modes experimentally. At the LHC, single top quarks will
be produced in ample numbers, allowing for detailed investigations.
However, the priority for the near future will be to establish a
single top-quark signal at ATLAS and CMS. Strategies for the
analysis of early LHC data have been devised by both 
collaborations~\cite{ATL-PHYS-PUB-2010-003,CMS-PAS-TOP-09-005}.

\Acknowledgements
The author acknowledges the financial support of the 
Helmholtz-Alliance {\it Physics at the Terascale}.

\end{document}